# Chapter 4
# MRI-guided HIFU Methods for the Ablation of Liver and Renal Cancers

**Baudouin Denis de Senneville, Chrit Moonen, Mario Ries**

**Abstract** MRI-guided High Intensity Focused Ultrasound (MRI-HIFU) is a promising method for the non-invasive ablation of pathological tissue in many organs, including mobile organs such as liver and kidney. The possibility to locally deposit thermal energy in a non-invasive way opens a path towards new therapeutic strategies with improved reliability and reduced associated trauma, leading to improved efficacy, reduced hospitalization and costs. Liver and kidney tumors represent a major health problem because not all patients are suitable for curative treatment with surgery. Currently, radio-frequency is the most used method for percutaneous ablation. The development of a completely non-invasive method based on MR guided high intensity focused ultrasound (HIFU) treatments is of particular interest due to the associated reduced burden for the patient, treatment related patient morbidity and complication rate. The objective of MR-guidance is hereby to control heat deposition with HIFU within the targeted pathological area, despite the physiological motion of these organs, in order to provide an effective treatment with a reduced duration and an increased level of patient safety. Regarding this, several technological challenges have to be addressed: Firstly, the anatomical location of both organs within the thoracic cage requires inter-costal ablation strategies, which preserve the therapeutic efficiency, but prevent undesired tissue damage to the ribs and the intercostal muscle. Secondly, both therapy guidance and energy deposition have to be rendered compatible with the continuous physiological motion of the abdomen.

## 4.1 Introduction

Hepatic and renal cancers account for 700 000 and 115 000 deaths per year in the word, respectively (Ferlay *et al* 2010). Worldwide, liver cancer is the fifth most frequent cancer in the male population and the seventh in the female. In East and Southeast Asia, and in West and Central Africa, primary liver cancer is the most common form of liver cancer (American Cancer Society 2011). Primary liver can-

B. Denis de Senneville (✉) • C. Moonen (✉) • M. Ries (✉)

Imaging Division, University Medical Center Utrecht, Heidelberglaan 100, P.O. Box 85500, 3508 GA Utrecht, Netherlands.

Mathematical Institut of Bordeaux (IMB), UMR 5251 CNRS/University of Bordeaux, 33400 Talence, France.

E-mail: b.desenneville@umcutrecht.nl ; c.moonen@umcutrecht.nl ; m.ries@umcutrecht.nl

cer is most frequently the result of a chronically damaged liver, with viral hepatitis, alcohol abuse and obesity as the predominant contributing factors (American Cancer Society 2011). To date, treatment options for primary liver cancer include liver transplantation, resection and ablation. However, only about 25% of primary liver cancer patients are eligible candidates for these curative therapies. The most frequent counter indications include the tumor size and location, underlying parenchymal disease or multi-focal lesions. This translated into a growing interest in mini-invasive local therapy, such as Radio Frequency-ablation (Lele *et al* 1967, Fry *et al* 1978), which the American Association for the Study of Liver Disease recommends for patients with less than three primary tumors with a diameter less than 3cm. For more advanced disease, transarterial chemo-embolization (TACE), radio-embolization and systemic chemotherapy are frequently offered as palliative measures.

In Europe and North America, metastatic liver tumors are the most common form of liver cancer (American Cancer Society 2011), whereby the majority of the primary tumors are located in the breast, lung, colon, prostate and rectum. In particular, colorectal cancer (CRC) has over 400000 cases per year, the second largest diagnosed cancer, of which about 70% develop metastatic disease in the liver (Ruers *et al* 2002). The occurrence of metastatic disease in the liver means that the primary cancer has reached stage IV and requires systemic therapy. Therapy for metastatic disease in the liver originating from CRC is flanking the curative therapy of the primary cancer. Similarly to primary liver cancer, the most frequently offered therapeutic approaches are surgical resection. However, similar to primary liver cancer, only about 20-25% of the metastatic disease is resectable, translating into a growing interest for mini-invasive local therapies, such as radiofrequency and laser induced ablation, cryotherapy or local embolization (Goldberg *et al* 2009). These are less limited with respect to patient selection.

Since local ablative therapy has become of increasing importance for both primary liver and metastatic liver cancer, , the possibility to target and ablate in a single session primary and metastatic cancer deep in the liver and kidneys in a non-invasive way has considerably amplified HIFU clinical interest in recent years.. Due to the associated reduced burden for the patient, treatment related patient morbidity and complication rate, HIFU represents a potential therapeutic alternative to patient that are currently not eligible for invasive or mini-invasive therapy.

From a historical point of view, the possibility to employ HIFU as an external energy source for non-invasive tissue ablation has been described by Lynn et al. as early as 1942 (Lynn *et al* 1942). Nevertheless, the first clinical experiments with HIFU were conducted in the field of neuro-surgery sixteen years later by Fry and coworkers (Fry *et al* 1958). One of the many technical limitations that hampered clinical adoption of this methodology at this point in time was the lack of a non-invasive method for interventional planning and guidance. As a consequence, the widespread introduction of ultrasound imaging systems with b-mode capabilities during the 1980's renewed the interest in this non-invasive therapeutic modality.

The feasibility of HIFU ablations with an extracorporeal device under ultrasound guidance in the liver was demonstrated by Vallancien and coworkers (Vallancien *et al* 1992). The first larger clinical studies were conducted by Wu and colleagues (Wu *et al* 1999, Wu *et al* 2004, Kennedy *et al* 2004) using ultrasound imaging with a JC ablation device from Chongqung Haifu (Chongqung, China) between 1997 and 2003.

In parallel, Cline *et al* had already suggested in 1992 the concept of MR-guided HIFU interventions (Cline *et al* 1992). This was followed up upon by several prototype HIFU systems that were fully integrated in whole body MRI systems (Cline *et al* 1995, Hynynen *et al* 1997, Jolesz *et al* 2002). Although these first systems were originally designed for MR-guided HIFU therapy of uterine fibroids, modifications to cope with the respiratory motion of the liver enabled first clinical studies using a modified ExAblate 2000 system (InSightec, Israel) by Okada *et al* and Kopelman *et al* (Okada *et al* 2006, Kopelman *et al* 2006).

With respect to the kidney, the first pre-clinical application of HIFU with an extracorporeal transducer on the kidney in an animal model was demonstrated by Chapelon and coworkers using ultrasound guidance (Chapelon *et al* 1992). Nevertheless, it took another thirteen years until the first clinical studies of HIFU ablations for renal masses under ultrasound guidance were reported in the same year by Häcker *et al* (Hacker *et al* 2006) and Illing *et al* (Illing *et al* 2005). More recent clinical studies explored the benefit of laparoscopic HIFU transducers (Klingler *et al* 2008) under ultrasound guidance.

MR-guided HIFU interventions in the kidney have so far been investigated only in preclinical experiments by Ries *et al* (Ries *et al* 2010) and Quesson *et al* (Quesson 2011). These carry the aim of demonstrating that MR-based real-time monitoring of temperature evolution and real-time motion compensation strategies, such as respiratory gating and beam steering, are feasible.

In particular the integration of HIFU-systems in clinical magnetic resonance imaging systems offers several compelling advantages (Okada *et al* 2006, Kopelman *et al* 2006). First, MRI allows acquisition of high-resolution 3D images with an anatomical contrast similar to diagnostic imaging, while the patient is in place on the therapeutic ablation system. This allows depicting both the lesion and organs at risk (OAR) directly before therapy, and thus takes directly into account recent physiological changes, such as tumor growth (or regression), and liver position modifications due to patient positioning on the ablation system. Furthermore, MRI allows non-invasive measurement of the local tissue temperature with a high precision and spatio-temporal resolution (for an extensive overview, see Denis de Senneville *et al* 2005, Rieke *et al* 2008). As a consequence, MR-Thermometry allows direct and continuous monitoring of energy delivery, and thus therapy progression (Hynynen *et al* 2006, Gellermann *et al* 2005, Denis de Senneville *et al* 2007). Furthermore, temporal evolution of temperature allows calculation of the thermal dose delivered (hereafter referred to as MR-dosimetry), which has been demonstrated as an empirical estimator of induced necrosis (Sapareto *et al* 1984). This is also applicable for the liver and kidney (Quesson *et al* 2011). Of particular importance here is the ability of monitoring not only the temperature in the target area, but also in the OAR and in the propagation part of the acoustic beam (Mougenot *et al* 2011, Baron *et al* 2013). Finally, dynamic contrast enhanced T1-weighted imaging (DCE-T$_1$wMRI) allows mapping the non-perfused volume (NPV) immediately after therapy, and thus validates the therapeutic endpoint on the fly.

In order to exploit these intriguing possibilities of MRI-guided HIFU for the ablation of liver and renal cancers, compared to other clinical HIFU applications two principal problems have to be overcome: The necessity to deposit the acoustic energy across the thoracic cage, and the necessity to render the targeting, energy deposition and interventional guidance compatible with the continuous physiological motion of the abdominal area.

## 4.2 Obstruction of the Ultrasonic Beam Path by the Thoracic Cage

As shown in **Fig. 4.1**, an important challenge for HIFU ablations in the liver and the kidney is obstruction of the acoustic beam path by the thoracic cage. For sonications in the cranial and lateral part of both organs, the ribs and the costal cartilage block part of the acoustic energy that is delivered by the ultrasound beam. This, in turn, leads to two principal problems:

Firstly, the high ultrasound absorption coefficient of the bone (Goss *et al* 1979) reduces the available acoustic beam power in the focal area. Moreover, the thoracic cage acts as an aberration, decreasing the focusing quality of the ultrasonic beam (Liu *et al* 2007, Bobkova *et al* 2010). Due to such beam aberrations, the required temperature increase in the target area may be significantly reduced (Liu *et al* 2005). This leads to either a significantly reduced volumetric ablation rate, or, if compensated for by an increased total emitted power of the transducer, to an increased risk of undesired near-field damage due to high acoustic power densities in the intercostal and/or abdominal muscle and ribs.

Secondly, the high acoustic absorption in the ribs and the cartilage can lead to local hyperthermia in the cortical bone, the bone marrow and the adjacent intercostal muscle. During energy delivery, temperature elevations in the ribs have been reported that were up to five times higher than the temperature in the intercostal space (Daum *et al* 1999). Such high local temperature elevations can lead to adverse effects, such as skin burns and intercostal tissue necrosis, which have been frequently reported in clinical studies (Wu *et al* 2004, Li *et al* 2007). Jung *et al* (Jung *et al* 2011) reported rib necrosis and diaphragmatic ruptures as the most frequent complications for HIFU treatments of hepatic tumors.

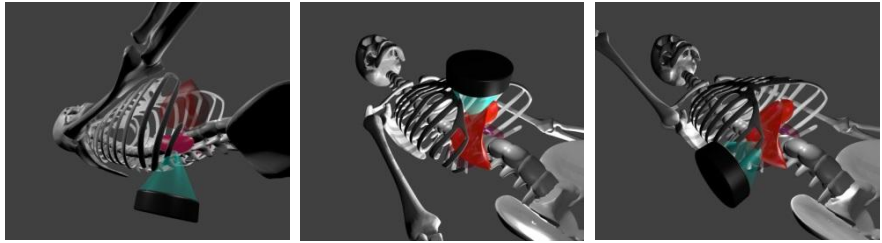

**Fig. 4.1 Obstruction of the acoustic beam path by the thoracic cage.** As the left illustration displays, only the caudal part of the kidneys can be reached by an extracorporeal transducer from dorsal direction, without parts of the propagation path being partially obstructed by the first and second floating rib. In particular the more cranially located left kidney is often only reachable by an intercostal sonication. With respect to the liver, only lesions in segments 4b and 5 can be reached by an extracorporeal transducer from a ventral position without any obstruction by the costal cartilage or the ribs, as shown in the center illustration. Therapy in segments 1, 2, 3, 4a requires angulated sonications. These partially interfere with the costal cartilage or the sternum. HIFU therapy in segments 6 and 7 of the liver generally require an intercostal sonication through the last two false ribs (9 and 10) from a lateral position, as shown in the right illustration. Otherwise, therapy in segment 8 is obstructed by both the false ribs and the costal cartilage (Image courtesy of Dr. Mario Ries, UMC Utrecht).

Several measures have been proposed to reduce or to avoid these adverse effects. In some cases, surgical removal of the part of the ribs that intersect with the acoustic propagation path has been applied to selectively perform HIFU treatment (Wu *et al* 2004). However, this negates the non-invasive nature of HIFU interventions and is likely to introduce new sources of adverse effects such as scar tissue, which in turn can lead to skin-burns.

The currently most commonly proposed method to mitigate excessive rib heating and restore focal point heating is beam shaping. Ebbini and Cain proposed in 1990 a HIFU transducer design in form of a sparse spherical phased array (Ibbini *et al* 1990). The main motivation of the use of spherical phased arrays is the combination of a geometric focus with the possibility to deflect the beam by modulating the phase of each individual transducer element. For intercostal sonications, this type of transducer design offers the additional possibility to selectively attenuate, or even deactivate, parts of the transducer surface. This has been demonstrated by McGough *et al* in 1996 (McGough *et al* 1996) and Botros *et al* in 1997 (Botros *et al* 1997) who proposed in theoretical design studies to adjust the amplitudes and phases of the voltage signals driving the individual phased array elements to limit rib exposure, while preserving the focal point intensity.

Civale and coworkers investigated in 2006 the use of a linearly segmented transducer for intercostal sonications (Civale *et al* 2006). He validated in both simulations and experiments the earlier prediction that the deactivation of edge segments leads to a significant reduction of the acoustic intensity on the ribs.

Since these early years, several refined methods to derive efficient beam shaping for intercostal sonications have been suggested. They fall into two families:

Approaches that use anatomical information in conjunction with acoustic simulations to derive the aperture function, and methods that directly use the therapeutic transducer for the detection of attenuating/scattering structures.

### 4.2.1 Apodization Methods based on an Anatomical Model of the Thoracic Cage using Binarized Apodization based on Geometric Ray-tracing

One of the first methods relying on anatomical information was proposed by Liu and coworkers (Liu *et al* 2007). Their method is based on geometric ray-tracing using CT images of the thoracic cage. In this approach, obstructed elements of a 2D phased array are identified by testing whether the normal vector of each element intersects with parts of the thoracic cage, as shown in **Fig. 4.2** on the left.

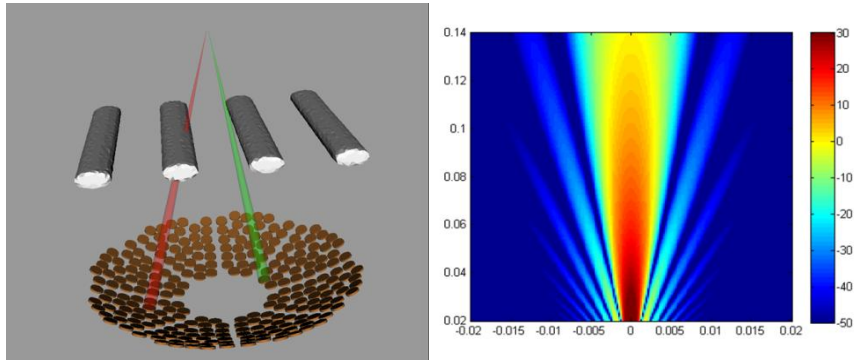

**Fig. 4.2 Binarized apodization approach for the compensation of the obstruction of the ultrasonic beam path by the thoracic cage.** The true acoustic emission profile of an individual cylindrical phased array element (shown in the right image) is approximated by a geometric ray in the normal direction of each element (shown on the left). Each ray is tested for intersections with scattering / attenuating anatomical structures, and obstructed elements (red) are selectively deactivated, while unobstructed elements (green) are amplified to partially compensate for the lost acoustic power (Image courtesy of Dr. Martijn de Greef, UMC Utrecht).

A subsequent selective deactivation of the obstructed elements (**Fig. 4.2**, red) and an amplification of unobstructed elements (**Fig. 4.2**, green) leads to a binarized apodization function, which can substantially reduce undesired heating of the ribs, while maintaining sufficient energy deposition at the target area. Quesson and coworkers demonstrated in 2010 the effectiveness of this approach both *ex-vivo* and *in-vivo* by using anatomical data derived from MR-images (Quesson *et al* 2010). The advantage of using MR-images in the context of MR-HIFU is that the anatomical information can be obtained with the patient in therapeutic position, thus taking local deformations and shifts of the organ vs. the thoracic cage directly into account. Furthermore, this method requires only a model of the thoracic cage with a moderate spatial resolution ($<1.5x1.5x1.5mm^3$), which can be obtained in a clinically relevant time frame. This approach is computationally efficient enough to be potentially feasible in a clinical workflow. The main drawback however is the coarse approximation of the true acoustic emission profile of an individual cylindrical phased array element (**Fig. 4.2**, right), which is in the form of a much more complex Bessel-function with the pointing vector of a plane wave. This neglects both the substantial off-axis energy emission of each element, which also contributes to undesired heating of the thoracic cage and diffractive effect of the ribs on the remaining acoustic field. Nevertheless, despite its limitations, this approach has been shown in several preclinical studies as quite effective (Bobkova *et al* 2010, de Greef *et al* 2014, Gélat *et al* 2014).

### 4.2.1.1 Phase Conjugation

Phase conjugation is a more advanced beam shaping method introduced by Aubry and coworkers (Aubry *et al* 2008), which in turn is based on the principle of time-reversal (Fink *et al* 1997, Fink *et al* 2003). In this approach, a point source is placed in the focus and the acoustic wave is propagated towards the transducer. Subsequently, the received signal in each individual transducer element is recorded. By emitting these recorded signals in a time-reversed fashion, a focus will be created at the target location under minimal exposure of the ribs, as most of the incident energy on the ribs will not be incident on the transducer. This approach relies on the linearity and the reciprocity of the wave equation in a non-dispersive medium. If both assumptions are fulfilled, the time-reversal process represents a spatio-temporally matched filter of the wave propagation operator (Tanter *et al* 2007). In its original implementation, this approach required a physical acoustic point source in the focus, and was therefore as an invasive technique clinically not feasible. However, if a high-resolution 3D representation of the tissue stack is combined with the appropriate acoustic impedance values of the tissues, the *phase conjugation* approach can be carried out virtually, i.e. in a simulation environment (Aubry *et al* 2008). Compared to binarized apodization, this approach performs a (full) phase-amplitude optimization for each transducer element. Furthermore, it has been shown to be among the most effective approaches to further reduce the energy exposure to the ribs, while maintaining the acoustic intensity in the focal point (Gélat *et al* 2014).

One of the drawbacks is that these methods are for computational reasons approximative and lengthy due to the fact the wave front is sampled with a sampling distance corresponding to several wavelengths, causing a mismatch between the forward and reverse acoustic field (Tanter *et al* 2001). Furthermore, this approach requires a large dynamic amplitude range of the transducer elements, which in-turn leads to a heterogeneous distribution of element power, thus local near-field overheating becomes an even larger risk, as well as overheating of individual transducer elements. In addition, this approach requires a complete and precise segmentation of the heterogeneous tissue stack between the HIFU transducer and target location. As a consequence, practical limitations with respect to the resolution and spatial fidelity of the acquired 3D model, and deviations from the assumed tissue properties will render the calculated solutions in practice suboptimal.

### 4.2.1.2 Constrained Optimization using the Boundary Element Method (BEM)

One of the limitations of binarized apodization based on geometric ray-tracing is that the effect of the apodization on the focus quality is not taken into account. This means that, although undesired heating of the thoracic cage is prevented, the impact of the element deactivation on the focal point amplitude might lead to a configuration that is therapeutically ineffective in the focus. This motivated Gélat and colleagues (Gélat *et al* 2014) to formulate the problem of focusing the field of a multi-element HIFU array inside the thoracic cage. This method ensured that the acoustic intensity on the ribs did not lead to undesired tissue damage as an inverse problem using the boundary element method (BEM) as the model capable of addressing the physical effects such as scattering and diffraction on 3D anatomical data (Gélat *et al* 2011). While a first preclinical study (Gélat *et al* 2014) suggests considerable potential of this approach, its current formulation is computationally extensive and requires a precise anatomical 3D model of each shot position for optimal results. Similar to the *phase conjugation* method, *constrained optimization* also requires a large transducer element dynamic amplitude range for it to be effective. As a consequence, future studies will have to investigate if this concept can be exploited for clinical applications.

### 4.2.1.3 Apodization Methods based on Direct Detection of Scattering or Attenuating Structures

One of the major drawbacks of intercostal apodization methods, which rely on an anatomical 3D model, is the requirement to rapidly and non-invasively map the anatomy between the energy source and the ablation area, and then to transform this anatomical data into a valid acoustic 3D model. If CT is chosen as the imaging modality on which the anatomical 3D model is based, then the anatomical 3D images are generally not acquired with the patient in the final therapeutic treatment position. Therefore, local anatomy deformations due to different patient positions and shifts of the organs vs. the thoracic cage that occur during transfer, limit the validity of the anatomic model. For MR-HIFU this can be omitted by using 3D MRI itself to derive an accurate spatial representation of the target anatomy and the scattering/attenuating structures. Nevertheless, this approach requires considerable image acquisition time, and a lengthy segmentation process that trans-forms the anatomical map into an acoustic model before the acoustic optimization of the transducer apodization can even begin.

Although each of the required steps is well understood, the ensemble is in practice often laborious, time consuming and error prone, requiring frequent user intervention for error correction and quality control. This complicates the clinical workflow of HIFU intervention. As a consequence, there has been a growing interest in developing methods that can directly detect scatterers and/or attenuating structures without an additional imaging modality and preferably without user intervention.

### *4.2.2 Decomposition of the Time-Reversal Operator*

One of the first approaches of this type for intercostal HIFU was suggested by Cochard and coworkers (Cochard *et al* 2009), firstly using a 1D linear phased array, and then demonstrated in 2013 using a sparse 2D array (Cochard *et al* 2011). The method was derived from the initial decomposition of the time-reversal operator (DORT) selective focusing method from Prada et al. (Prada 2002). In its original implementation, the DORT method was developed for adaptive focusing of ultrasonic arrays on strong back-scatterers. DORT relies on the acquisition of the backscatter matrix (*i.e.* the columns of this matrix relate the scattered signal received by all transducer elements from an excitation event of each transducer element). A singular value decomposition of this matrix leads to a set of eigenvectors that represent amplitude/phase combinations, which focus the beam on the individual scattering structures. Since the corresponding eigenvalues to the obtained eigenvectors allow their classification by their scattering magnitude, a linear combination of the eigenvectors can be computed. This combination represents a phase/amplitude combination for each transducer element that focuses the beam on several selected scattering structures simultaneously. For optimal intercostal HIFU, Cochard et al proposed to reverse this principle and to compose an amplitude/phase vector from the eigenvectors with the lowest eigenvalues, i.e. eigenvectors that focus the beam on acoustically transparent (*i.e.* unobstructed) areas in the beam path (Cochard *et al* 2009).

The main advantages of this approach are that it uses the HIFU transducer itself as the detector and allows performing the required acquisitions and calculations in not only a very short time frame, but also non-invasively. This allows potential adaptation of the apodization of the phased array elements for optimal intercostal sonications, even under the limitations of a clinically feasible and repetitive workflow for every different transducer position on the fly.

The main disadvantage of the DORT method is that it relies on the presence of strong scatterers and works best with well-resolved point-like scatterers. Although cortical bone represents such a strong scatterer, the complex shape of the ribs is not well represented by a low number of point sources. Furthermore, as **Fig. 4.3** il-lustrates, the ultrasonic propagation path is also frequently obstructed by attenuating structures, such as costal cartilage, which are not well suited for this detection method.

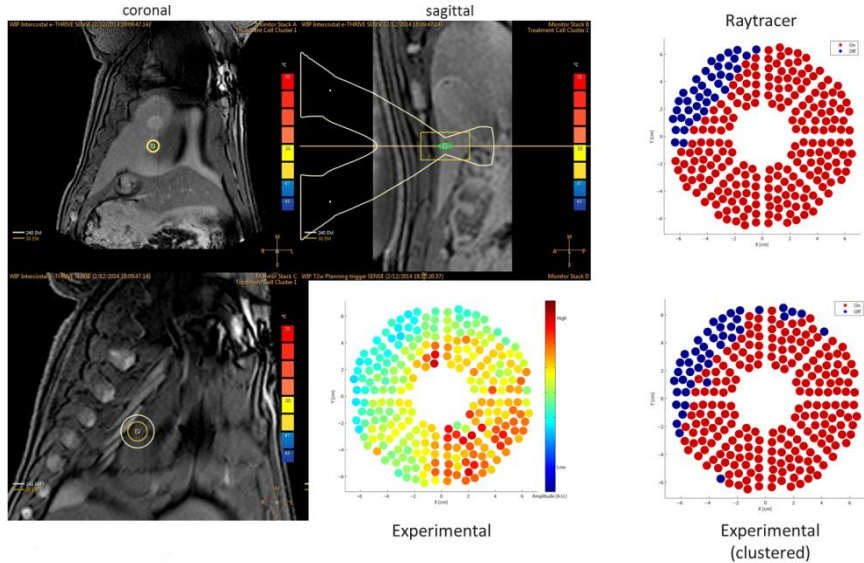

**Fig. 4.3 Automatic detection of beam obstruction by the thoracic cage during liver ablation in an *in-vivo* experiment.** The focus is placed in segment 4 of a porcine liver, as shown in the coronal (top left) and the sagittal (top center) $T_1$ weighted image. As indicated in the coronal image on the level of the sternum (bottom left), the beam cone intersects with parts of the costal cartilage. In this example, the binarized apodization based on geometric ray-tracing (top right) is based on a semi-automatic segmentation of a 3D $T_1$–weighted MRI of the costal cartilage,  resulting in a selective deactivation of obstructed transducer elements. The binarized apodization based on cavitation enhanced back projection (bottom right) leads to comparable results, but requires only a fraction of the acquisition and processing time (<1s vs. 5.3min). In addition, this approach provides an estimate of the relative attenuation between focus and transducer for each element (bottom middle) (Image courtesy of Pascal Ramaekers, UMC Utrecht).

As a result, a comparison of the DORT method in simulated experiments with other approaches (binarized apodization based on geometric ray- tracing, phase conjugation and constrained optimization as previously proposed (Gélat *et al* 2012) (see below)), indicated that although the DORT method results in an apodization which spares the thoracic cage from undesired acoustic intensity, this is more at the expense of the focal point pressure. In summary, although the DORT approach is conceptually very promising, *in-vivo* studies that validate the efficiency of the approach are to date still pending.

### 4.2.2.1 Pulse-Echo Detection

A much simpler approach was suggested by Marquet and colleagues (Marquet *et al* 2011)that exploits the A-mode imaging capabilities of the HIFU transducer directly in order to detect obstructed parts of the transducer. Similar to DORT, this approach exploits the strong backscattering of an emitted acoustic pulse from the thoracic cage. Here, the transducer channels are ordered according to the amplitude strength of their back reflected signal and clustered into obstructed and unobstructed elements. This approach has the advantage to be both very fast and comparably simple, thus having the potential to be compatible with a clinical workflow. The main drawbacks are that although cortical bone represents a strong acoustic reflector, cartilage signal reflection is much harder to differentiate from reflections from the adipose-muscle and muscle-liver/kidney tissue boundaries. This is further complicated by the fact that most HIFU transducers are designed as narrow-band systems at lower frequencies (0.75-1.5 MHz), thus being limited in their A-mode imaging quality.

### 4.2.2.2 Cavitation-enhanced Back-Projection

The original implementation of the *phase conjugation approach* requires a point source in the focus, which emits a spherical acoustic wave that is subsequently received by all transducer elements. As Tanter *et al* have shown, the time-reversal of this received amplitude/phase vector represents a spatially and temporally matched filter of the propagation operator through the heterogeneous medium (Tanter *et al* 2001). The drawback of this approach is that it is invasive, and thus for clinical applications unfeasible. The aim of work for intercostal HIFU with a transducer apodization based on time-reversal was aimed to circumvent this limitation: Aubry *et al* (Aubry *et al* 2008) virtualized the physical time-reversal measurement with model based solutions (**see section 4.2.1**), while Cochard and coworkers (Cochard *et al* 2009) recorded the complementary information, i.e. the backscatter from the ribs.

The key idea of *cavitation-enhanced back-projection* is to replace the time-reversal experiment (i.e. using the propagation of a spherical wave from the focus to the transducer in order to derive the transmit apodization of the transducer) with a true pulse echo experiment between the transducer and a point scatterer in the focus. However, this requires placing a sufficiently large point scatterer in the focus of the transducer in a non-invasive way. This is achieved by emitting a first short pulse of ultrasonic energy through all transducer elements simultaneously, which are used to create a sufficient peak negative pressure in the focus to induce non-inertial cavitation. The resulting bubble cloud in the focal area represents a spatially confined cloud of point scatterer. As a result, consecutive ultrasonic waves are then reflected by the cavitation bubbles back onto the transducer.

The relative signal strength received from the back-reflected wave by each transducer element represents a measurement of the relative attenuation between the focus and each individual transducer element. In order to suppress undesired echoes other than those originating from the cavitation bubble cloud, a pulse inversion sequence was used. Using such a sequence, the majority of the received signal originates from the reflections of the cavitation bubbles.

This proposed method allows rapid mapping of any aberrating structures in each of the individual beam paths of the transducer elements before a HIFU sonication is started, and consecutively calculates the appropriate apodization law. Similar to the invasive *phase conjugation* method, this approach takes both absorbing and scatterering structures in the beam into account, whereby the acquisition and processing times are comparable with the DORT approach (<1s), and thus entirely compatible with the clinical workflow.

The principal disadvantage of *cavitation-enhanced back-projection* is the requirement to induce stable cavitation. Stable cavitation requires, in particular in deeper tissue layers, transducer and generator systems which can deliver large peak acoustic pressures. While this increases the complexity and the cost of the HIFU system, it also increases the risks of adverse effects. Power control of nonlinear energy deposition is challenging and pulses in the 3-6.5MPa pressure regime also increases the probability of potentially harmful inertial cavitation events, which in turn can lead to undesired tissue damage (Hwang *et al* 2006, Miller *et al* 2007).

## 4.3 Challenges associated with Physiological Motion of the Liver and Kidney

Besides the necessity to deposit the acoustic energy across the thoracic cage, the second major challenge for non-invasive HIFU therapy in liver and kidney, with respect to both energy delivery and therapy guidance, is physiological motion. It is therefore important to differentiate between the sources and the time-scale of the different types of physiological motion and the adequate measures in more details:

*Respiratory motion:* The liver and the kidney of an adult patient move under free-breathing conditions with a periodicity of around 3–5 s, and a motion amplitude of 10–20 mm, as shown in **Fig. 4.4**. While the motion pattern of free-breathing patients is over longer episodes (<30-45s) periodic, it is frequently subject to changes in amplitude, phase and frequency before a new stable breathing rhythm is reached. In particular, the occurrence of involuntary spontaneous motion events, such as swallowing, coughing or muscle spasms, is hard to predict and interrupts a regular breathing pattern.

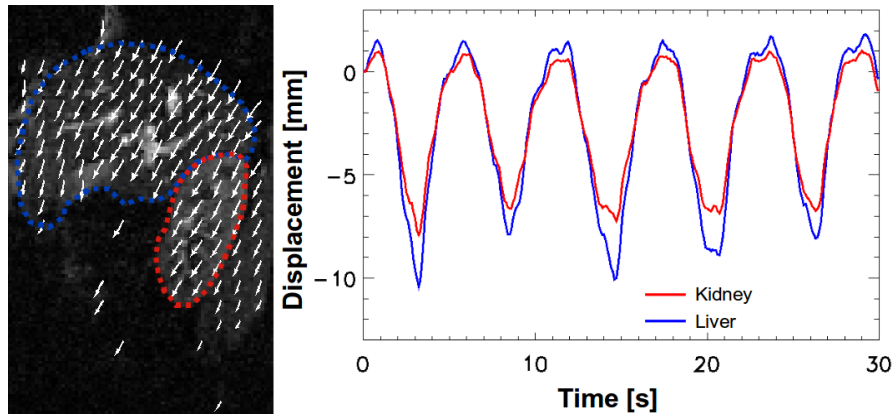

**Fig. 4.4 Proof of concept of the motion estimation process performed on-line in the abdomen of a healthy volunteer for liver (inside blue contour) and kidney (inside red contour).** The estimation of organ motion is performed using real-time optical-flow algorithm as described in (Denis de Senneville *et al* 2011) with the sagittal anatomical images reported on the left. The vertical component of the estimated motion is reported on the right in the liver and the kidney (Image courtesy of Baudouin Denis de Senneville, CNRS/UMC Utrecht).

As a consequence, there is a growing trend to perform non- and minimally invasive therapy under deep sedation. Deep sedation is induced by an intravenous infusion of a hypnotic/amnestic agent, such as propofol or sodium thiopental, to reduce the probability of involuntary spontaneous motion events.

The additional use of analgesic drugs, in particular analgesics based on opioids, lead to respiratory depression, sometimes used to reduce in addition both the respiratory frequency and amplitude. Although the patient still displays spontaneous respiration, this measure can significantly increase the time fraction during which the abdomen is not subject to respiratory motion.

The next step from this intermediate regime is to induce complete respiratory depression and to control the respiratory cycle by an external mechanical ventilator. This approach leads to repetitive and stable respiratory motion over long durations, whereby the amplitude and the frequency can be adjusted within the boundaries of sufficient blood-oxygen saturation according to the required interventional workflow.

***Peristaltic motion:*** A second important source of physiological motion is induced by peristaltic and digestive activity in the digestive tract. Although the time-scale of peristaltic motion events depends on the particular source, such as bladder filling with urine, the development of digestive gases or the passage of digestive products in the gastrointestinal tract, the resulting abdominal organ position shifts usually occur on a scale of several minutes (Mirabell *et al* 1998, Langen *et al* 2008). Although peristaltic motion is with respect to both amplitude and speed a magnitude below respiratory induced displacements, it is generally non-reversible and a-periodic. Peristaltic motion can clinically be moderated by several measures: Peristaltic bowel motion and the development of peristaltic gases can be reduced by adjusting the diet of the patient prior to the intervention (Smitmans *et al* 2008). Similarly, administration of Butylscopolamine, which represents a peripherally acting antimuscarinic and anticholinergic agent used as an abdominal-specific antispasmodic (Emmott *et al* 2008). Furthermore, shifts in the lower abdomen due to bladder filling can be reduced by the use of Foley catheters (Mira-bell *et al* 1998).

***Spontaneous motion:*** Finally, spontaneous motion is considered one of the most challenging types of physiological motion since it occurs infrequently, on a very short time-scale and is in general irreversible. It is particularly problematic for long interventions that require the patient to remain in an uncomfortable position. In the past, this problem has been alleviated in the field of external beam therapy by using either restraints, such as molds or casts (Verhey *et al* 1995), by sedating the patient (Zhang *et al* 2010), or by introducing general anesthesia (GA).

### 4.3.1 Motion Compensation Strategies for HIFU Ablation on Abdominal Organs

As shown in **Fig 4.5** the typical workflow of an MRI-guided HIFU intervention starts with an initial planning (5-15min), followed with a sequential set of energy depositions (5-60s each) together with the associated cool down delay (30-180s, depending on the energy density in the near-field).

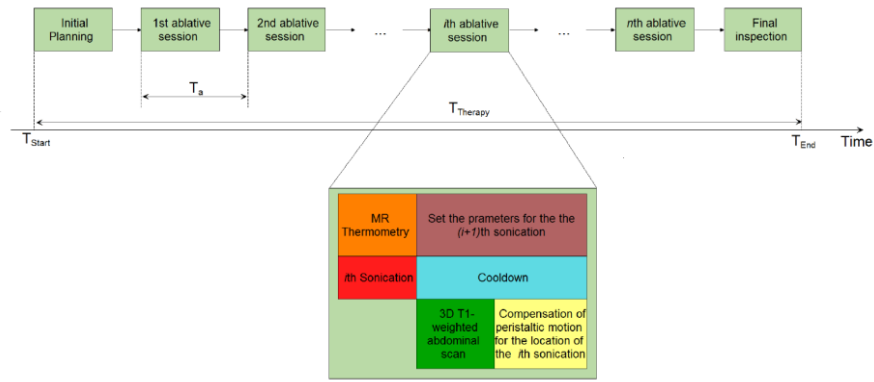

**Fig. 4.5 Typical workflow of an MR-guided HIFU intervention.** During the initial planning a set of anatomical images is obtained on which the planning target volume (PTV) around the tumor is delineated. In general, the PTV is ablated in smaller subvolumes, which allows intermediate tissue layers to cool down during the intervention. At the end of the intervention a dynamic contrast enhanced 3D MRI validates the therapeutic endpoint (Image courtesy of Cornel Zachiu, UMC Utrecht).

This therapeutic phase can have duration of up to three hours. It is concluded with a set of physiological and functional MRI datasets which validate the therapeutic endpoint. As **Fig 4.5** displays, while peristaltic motion is during the short duration of each of these episodes not a major problem, these shifts can become challenging when considered at the time scale of the entire intervention.

The influence of respiratory motion during the initial and the final phases of the intervention are generally addressed with the established measures of diagnostic MRI: Either respiratory gating or breath holding. During the therapeutic phase, both for MR-guidance and energy delivery, respiratory induced motion needs to be addressed individually. Several approaches have been investigated to achieve this:

*Induced Apneas:* The most simple and efficient way to prevent undesired respiratory motion of abdominal organs is to temporary interrupt the respiratory cycle. This is generally achieved by inducing general anesthesia with hypnotic/amnestic agents, which are given in conjunction with analgesics based on opioids to achieve a full respiratory depression. This allows controlling the respiratory cycle entirely by mechanical ventilation, which in turn can be synchronized with the therapeutic energy delivery. Several clinical studies have demonstrated the feasibility of this approach: Both Gedroyc *et al* (Gedroyc *et al* 2006) and Kopelman *et al* (Kopelman *et al* 2006) have successfully ablated Hepatic tumors by repetitive induced apneas. The main advantage of energy delivery during induced apneas is that this approach is compatible with any type of clinical HIFU equipment and does not require major modifications with respect to beam steering capabilities or beam amplitude modulation capabilities. The main drawback of this approach is that it reduces the non-invasive nature of MR-guided focused ultrasound interventions. While for primary tumor therapy of patients with a good general condition, an intervention under sustained general anesthesia of 2-3h is generally considered clinically acceptable, the situation is more complicated when metastatic disease or patients in a poor general state are considered. In particular, for the case of metastatic disease, where local HIFU therapy represents only one aspect out of a combination of systemic and local therapy measures, an increased invasiveness of HIFU therapy is likely to reduce both the therapeutic possibilities and patient eligibility.

*Gating Strategies:* As an alternative to induced apneas, respiratory-gated energy delivery strategies also allow the addressal of respiratory motion. As shown in **Fig. 4.4**, the respiratory cycle of an adult patient under resting conditions displays typically a time-window of 1-3s when the diaphragm remains stationary. Therefore, a time synchronous amplitude modulation of a spatially stationary HIFU beam allows depositing the entire acoustic energy in a precise location despite the periodic displacements of the target. However, the required HIFU amplitude modulation leads to a significant reduction of the duty cycle of the ablation process. Since both kidney and liver display a high perfusion rate and consequently heat evacuation, a respiratory-gated energy delivery significantly reduces the achieva-ble temperatures for larger sonication volumes (Cornelis *et al* 2011). The situation is further complicated by the tendency of the respiratory cycle to be under free-breathing, only periodic/stable for shorter durations (10-30s), and by slow drifts of the diaphragm resting position of 8-9mm/h..

A gated energy delivery is considerably more favorable for patients under full anesthesia, where analgesic drugs allow the induction of full respiratory depression, and both breathing frequency and volume are adjusted by an external mechanical ventilator. As shown in **Fig 4.6,** mechanical ventilation allows the maintenance of an entirely periodic breathing pattern that allows gated HIFU energy delivery with a duty cycle of up to 80% for extended durations of several hours.

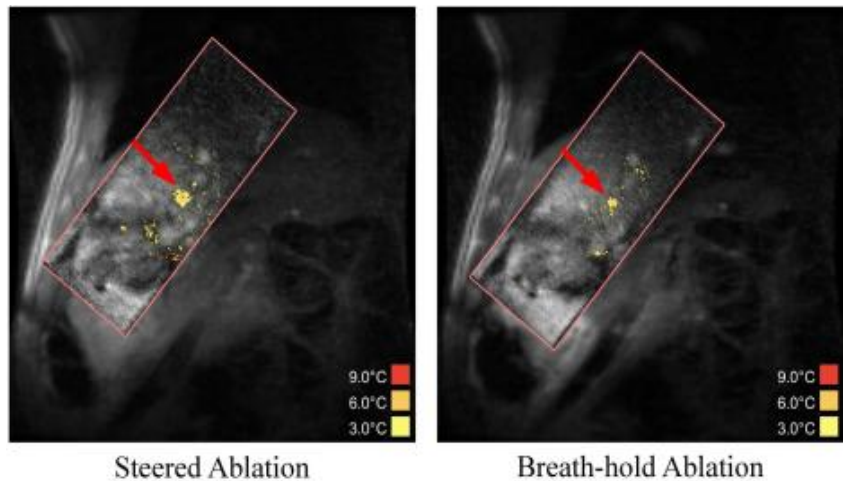

Steered Ablation                     Breath-hold Ablation

**Fig. 4.6** *In-vivo* **HIFU steered and breath-hold ablations.** These two images show representative images of a steered (left) and breath-hold (right) ablation *in-vivo* at the time point when the maximum temperature reached 10°C above baseline. The steered ablation took 31.2 s to reach this point, whereas the breath-hold ablation took 26.7 s. The steered ablation required 16.7% more energy than the breath-hold case. Reproduced with permission from (Holbrook *et al* 2014).

The clinical feasibility of this approach has been demonstrated by Okada and coworkers (Okada *et al* 2006) for the treatment of hepatocellular carcinoma. More recent preclinical work introduced several methodological improvements, such as MR pencil-beam navigators (Wijlemans *et al* 2014a) and optical tracking as the gating source (Auboiroux *et al* 2014), and additional motion correction strategies to account for peristalsis and slow target drifts.

As a consequence, for patients under full anesthesia, respiratory gating represents an interesting alternative to induced apnea, allowing a more continuous workflow compared to the cyclic apnea/re-oxygenation workflow of the later. In particular, recent work (Wijlemans *et al* 2014b) to evaluate the compatibility of respiratory-gated sonications under Procedural Sedation and Analgesia (PSA), in-stead of general anesthesia, is likely to reduce the risk of complications and to shorten patient recovery.

**Beam Steering Strategies:** The most promising approach to render ablations under free-breathing conditions compatible with an efficient energy delivery is to exploit the beam steering capabilities of modern phased array-transducers to continuously reposition the focus on the current position of the target. Contrary to amplitude modulated delivery schemes, such as gating, beam steering allows the continuous deposition of acoustic energy over periods of time that exceed the duration of the respiratory cycle. This is of particular importance in tissues which display high perfusion rates, such as the kidney and liver (typical liver perfusion 65-100 mL/min/100mL, kidney perfusion 287-379 mL/min/100mg (Koh *et al* 2008)).

*MR-based tracking:* In the context of real-time guidance of hyperthermia, MR-navigator echoes have initially been proposed to provide motion information in combination with MR images (de Zwart *et al* 2001). These techniques are however inadequate to estimate complex organ deformation because the motion estimate is generally restricted to translational displacement. Since modern MRI acquisition methods now allow the rapid acquisition of large data volumes, combined with an excellent tissue contrast and high spatial resolution, image-registration techniques (Sotiras *et al* 2013) were recently employed to estimate on-line organ displacements from anatomical images. Complex deformations can for example be estimated on a voxel-by-voxel basis using optical flow based approaches (Barron *et al* 1994), which assume a conservation of local pixel intensities along the target displacement. The potential of such techniques for real-time MR-guided HIFU on mobile targets was first demonstrated in 2007 (Denis de Senneville *et al* 2007). However, this approach faces the following challenges:

First, both the update rate and latency of measured displacement information may hamper steering accuracy. While the image sampling frequency is limited by the MR-acquisition time, the latency is determined by the remaining acquisition time after echo-formation, the image processing time, the switching time of the HIFU-generator and the required data transport. For typical abdominal organ motion, the delay between the actual time of displacement and when motion information is available must remain below 100 ms (Ries *et al* 2010). The image registration task is however highly computationally intensive and recent approaches take benefit of a combined CPU/GPU architecture by offloading computational intensive calculations to the GPU, thus freeing the CPU for pipeline management and data preparation. That way, the image registration step can be dedicated to the GPU, which allows completing all calculations on a typical image size of 128×128 voxels in about 100 ms, as shown in (Roujol *et al* 2010, Roujol *et al* 2011).

Secondly, optical-flow based algorithms rely on the assumption of conservation of local intensity along the trajectory which can be violated during thermotherapy because rapid MR-imaging is in general associated with low Signal-to-Noise Ratio SNR. In addition, since the tissue is heated, several MR relevant tissue properties; such as $T_1$, $T_2$ and $T_2^*$ relaxation times, are subject to change during imaging (Gra-ham *et al* 1998). This leads to local intensity variations, which in turn can be mis-interpreted by optical-flow based algorithms as "false motion". A Principal Component Analysis (PCA) based motion descriptor was thus recently introduced in order to characterize in real-time complex organ deformation (Denis de Senneville *et al* 2011) (Denis de Senneville *et al* 2015) . PCA was used to detect, in a preparative learning step, spatio-temporal coherences in the motion of the targeted organ. Then, during hyperthermia, incoherent motion patterns could be discarded. This method allowed maintenance of only the physiological components of estimated motion in the registration process, permitting reduction in the noise of the estimated displacement. In addition, the PCA-based motion descriptor provides a flow field that is consistent with the learned model. It is also robust under the assumption of global brightness constancy, but allows local intensity variations.

Finally, although it is well established that MR-imaging can provide motion estimates with a high spatial resolution, it is difficult in practice to acquire on-line 3D isotropic images due to technical limitations, spatial and temporal resolution trade-offs and low Signal-To-Noise Ratio (SNR) associated with fast 3D acquisition sequences. Real-time target tracking of abdominal organs depends on high frame-rate imaging; therefore it is not compatible with methods, such as respiratory gating or the acquisition of extended 3D volumes. In practice this limits MR-imaging to the acquisition of 1-3 slices, with modest spatial resolution in the slice direction. One approach consists of aligning the normal vector of the slice orthogonal to the motion vector, and thus to contain the entire motion cycle within a 2D imaging slice, as suggested in one report(Denis de Senneville *et al* 2007). This, however, imposes severe constraints on the imaging geometry, which might be for anatomical or diagnostic reasons unfavorable. It is often not possible to ensure that the target area remains observable during the entire motion cycle by a single static image slice. Furthermore, although the motion trajectory of the kidney and the lower part of the liver can be approximated by a linear shift, the true trajectory has a shape in 3D space. In particular, the upper part of the liver, which is subject to an elastic deformation, is hard to contain in a static 2D imaging slice during the entire respiratory cycle. Since extensive 3D-volume imaging is in practice hard to achieve with sub-second resolution, alternative approaches that dynamically adapt the image location to the current target location, have been proposed as a possible solution. The slice position is hereby continuously adjusted to the current target location using fast pencil beam navigator echos (Ries *et al* 2010, Köhler *et al* 2011) or ultrasound echoes (Gunther *et al* 2004, Feindberg *et al* 2010). In addition, a real-time image-based motion estimation algorithm applied on the image stream allows obtaining the in-plane target position sub-voxel precision. This, in combination with the retained slice tracking position, may describe the complete target position in 3D space, which can be conveniently employed to adjust the beam position (Ries *et al* 2010). Alternative strategies are under active investigation that consists in estimating the 3D motion from 2D real-time MRI, using for example one or several volumetric scans obtained before the intervention (Arnold *et al* 2011, Brix *et al* 2014, Stemkens *et al* 2014). These techniques however must still be validated in combination with MR-HIFU.

*US-based tracking:* The simultaneous use of US imaging and HIFU is difficult because of their mutual interferences. However, the combination of the two devices represents a promising approach to operate thermometry/dosimetry and beam steering tasks, independently and with different temporal resolutions. The acquisition of US signal and HIFU sonication must be interleaved. In the context of real-time guidance of hyperthermia, ultrasonic echoes have been used in the past to provide motion information with high temporal resolution in conjunction with MR images. Pernot and coworkers (Pernot *et al* 2003) used 4 pulse receivers to estimate the 3D displacement of the targeted organ (3 transducers were required to estimate the displacement, and another one was added to increase the robustness of the process). In the context of real-time guidance of hyperthermia, ultrasonic echoes have been used in the past to provide motion information with high temporal resolution in conjunction with MR images in (Lourenço da Oliveira *et al* 2010). However, using these techniques the estimated motion information is restricted to knowledge outside the heated zone because of the echo perturbation induced by the temperature rises with ultrasonic echoes. In addition, the challenge of US echoes increases when US waves are obstructed by ribs and/or air in the beam path. Consequently, it has been recently proposed to use 2D ultrasound echography as an additional imaging modality for continuous target tracking under MR-HIFU environment (Gunther *et al* 2004, Feinberg *et al* 2010, Ries *et al* 2012, Auboiroux *et al* 2012, Denis de Senneville *et al* 2014).

## 4.3.2 MR guided Thermometry and Dosimetry in Abdominal Organs

One of the most intriguing possibilities of MRI-HIFU is the possibility to real-time monitor temperature changes during an HIFU sonication non-invasively. Initial approaches focused on the temperature dependence of several MR observable tissue properties (such as $T_1$ and $T_2$ relaxation time of water protons, molecular diffusion constant of water…), which have been intensively studied in the literature. The associated thermometric MR-methods are reviewed elsewhere (Denis de Senneville *et al* 2005, Rieke *et al* 2008). The most promising candidate is real-time MR-thermometry based on the water proton resonance frequency (PRF), mainly because of its near-independence on tissue composition. Tissue necrosis can also be empirically estimated during the interventional procedure using the calculation of the accumulated thermal dose based on dynamic MR thermometry (Sapareto *et al* 1984). MR-thermometry/dosimetry is thus a promising candidate to assess an on-line retroactive control of therapy and ensure an accurate thermal energy delivery with HIFU.

Nevertheless, artifacts in the temperature measurements have a large impact on the precision of the thermal dose estimate due to its exponential dependence on the temperature, and will accumulate during the intervention due to the integration. A first challenge arises because the MR-thermometry based on the PRF shift of moving targets, such as the abdominal organs, is complicated by the continuous target motion through an inhomogeneous and time-variant magnetic field (Peters *et al* 2000). Respiration or cardiac-induced organ displacement and deformation will modify the local demagnetization field, and thus also the local magnetic field, experienced by the target organ. This will in turns result in temperature artifacts (De Poorter *et al* 1994, Young *et al* 1996). A second challenge rises from the fact that MR-thermometry is limited by the available signal-to-noise ratio (SNR). An inherent trade-off exists between spatial resolution, volume coverage and scan time. It is therefore currently difficult to acquire in real-time 3D isotropic thermal maps of a large field-of-view.

### 4.3.2.1 Compensation of Motion related Errors in Thermal Maps

Several strategies have been proposed to reduce the impact of breathing activity on MR thermometry in abdominal organs:

***Gated strategy:*** Using respiratory-gated strategies, MR acquisition is synchronized to a stable period of breathing activity. To this end, a temporal window of 1 second is employed during the exhalation phase of the respiratory cycle. The breathing motion pattern can be assessed on-line using various types of qualitative sensors, such as breathing belt (Moricawa *et al* 2004) or quantitative surrogates, such as MR navigators (Vigen *et al* 2003). This way, motion artifacts on thermal maps can be reduced, limiting however the temporal resolution to the respiratory frequency (Weidensteiner *et al* 2004).

***Non-gated strategy:*** Non-gated strategies have been recently proposed to provide continuous and regular temperature updates with a high temporal frequency during the entire respiratory cycle. However, precise modeling of the inhomogeneous magnetic field *in-vivo* and under real-time conditions is difficult to achieve, and consequently several alternative simplified strategies have been proposed to allow correcting motion related errors in PRF-based MR thermometry. For this purpose, two approaches emerged as promising candidates to enable continuous MR-thermometry of abdominal organs under free-breathing conditions: "Referenced" and "Referenceless" MR thermometry.

Referenced MR thermometry consists of analyzing phase artifacts with motion during a pre-treatment step performed prior to the intervention. A reference dataset of magnitude and phase images is recorded during the motion cycle of the organ. For this intention, the motion cycle must be sampled with a sufficient density in order to avoid discretization errors. With typical imaging frame-rates of 5-10 images per second, and a respiration frequency of 3-5 s, this pre-treatment step can be completed in a relatively short duration of 10-20 s. Phase artifacts, due to the periodical motion of the respiration cycle, are then addressed either by a multi-baseline strategy (Vigen *et al* 2003, Denis de Senneville *et al* 2007, Quesson *et al* 2011), or alternatively by applying a phase correction based on a model of the phase variation in dependence of the estimated target motion (Hey *et al* 2009, Denis de Senneville *et al* 2011).

Using referenceless MR thermometry, a background phase is estimated by fitting a polynomial function to the measured phase obtained from a region of interest (ROI) outside the treatment area, which is assumed to remain at body temperature as described by (Rieke *et al* 2004, McDannold *et al* 2008, Holbrook *et al* 2010). Recent updates of this approach aim at avoiding fitting problems due to spatial phase wraps, as well as determining the appropriate size and location of the ROI, and the optimal polynomial order for the phase fit can be determined before heating (Kuroda *et al* 2006, Zou *et al* 2013, Caiyun *et al* 2014). In particular, the computation of background phase using dipole-based filtering (Liu *et al* 2011) or harmonic interpolation (Schweser *et al* 2010, Salomir *et al* 2012) have also been investigated to further improve the performance of this approach.

A hybrid of both the multi-baseline and the referenceless methods has also been investigated: A direct combination of both techniques to compute each temperature map has been proposed by (Grissom *et al* 2010), as well as a temporal switch method (Denis de Senneville *et al* 2010). The latter initially employs the multi-baseline algorithm to continuously provide temperature maps across the entire field of view. In case of spontaneous movement during the intervention, for which no reference phase is pre-recorded, the correction strategy is updated dynamically from "referenced" to "referenceless" MR-Thermometry.

### 4.3.2.2 Challenges Associated with Real-time Volumetric MR-Temperature Imaging

Finally, 3D isotropic MR-Thermometry in a large field-of-view is a prerequisite to achieve an accurate monitoring of the thermal dose measured in the targeted area, as well as the collateral damages (such as edema induced in the near field close to the skin or due to heating of the bone by absorption of the acoustic energy). Several MR acquisition/reconstruction techniques have been employed to reduce the scan time of PRF sequences (Tsao *et al* 2003). In a study by Quesson *et al* 2011, volumetric MRI thermometry in pig livers was achieved with an update rate of 400 ms on a volume of five slices, together with a relatively high temperature precision of 2 °C. More recently, post-processing techniques have been also introduced to compute 3D MR-temperature maps in real-time using sparse sampling or Kalman filtering approaches (Todd *et al* 2013, Roujol *et al* 2012, Denis de Senneville *et al* 2013).

## 4.4 Conclusion

Since the introduction of the concept of MRI-guided HIFU for interventional oncology twenty years ago, a considerable interest has been directed towards ablation methods for liver and renal malignancies. In particular in the last decade, extensive preclinical research has meanwhile identified several clinically feasible solutions for the most challenging problems and several clinical pilot trial studies have been conducted.

Nevertheless, the substantial progress in both MR-guidance and refined acoustic energy delivery has so far not led to widespread clinical acceptance. The main reason is a lack of dedicated MRI-HIFU equipment, which must still be optimized for liver and kidney cancer therapy. The currently employed equipment and its associated methodology leads to many compromises, such as low volumetric ablation rates, or requirements such as general anesthesia, which renders this approach on par with other already clinically established mini-invasive techniques, such as RF-ablation.

However, recently  we have seen a rapid development of cancer screening modalities, which in turn has meanwhile resulted in an increasing number of patients where the primary tumor is detected at an earlier stage. As a consequence, the demand for non-invasive local therapy with non-ionizing radiation is likely to increase in the future.

## 4.5 References

American Cancer Society, Global cancer facts and figures: 2nd edition, Atlanta: American Cancer Society; 2011.

Ferlay J, Shin HR, Bray F, Forman D, Mathers C and Parkin DM (2010) Estimates of worldwide burden of cancer in 2008. Int J Cancer, 127:2893–2917.

Lele PP (1967) Production of deep focal lesions by focused ultrasound-current status. Ultrasonics, 5:105–112.

Fry FJ and Johnson LK (1978) Tumor irradiation with intense ultrasound. Ultrasound Med Biol, 4:337-341.

Ruers T and Bleichrodt RP (2003) Treatment of liver metastases, an update on the possibilities and results. Eur J Cancer, 238:1023–1033.

Goldberg SN, Grassi CJ, Cardella JF, Charboneau JW, Dodd GD III, Dupuy DE, Gervais DA, Gillams AR, Kane RA, Lee FT Jr, Livraghi T, McGahan J, Phillips DA, Rhim H, Silverman SG, Solbiati L, Vogl TJ, Wood BJ, Vedantham S and Sacks D (2009) Image-guided tumor ablation: standardization of terminology and reporting criteria. J Vasc Interv Radiol, 20:S377–S390.

Lynn JG, Zwemer RL, Chick AJ and Miller A (1942) A new method for the generation and use of focused ultrasound in experimental biology. J Gen Physiol, 26:179-193.

Fry FJ (1958) Precision high intensity focusing ultrasonic machines for surgery, Am J Phys Med, 37:152-156.

Vallancien G, Harouni M, Veillon B, Mombet A, Praponitch D, Brisset JM and Bougaran J (1992) Focused extracorporeal pyrotherapy: feasibility study in man. J Endourol, 6:173–181.

Wu F, Chen WZ and Bai J (1999) Effect of high-intensity focused ultrasound on the patients with hepatocellular carcinoma: preliminary report. Chin J Ultrasonogr, 8:213–216.

Wu F, Wang ZB, Chen WZ, Zhu H, Bai J, Zou JZ, Li KQ, Jin CB, Xie FL and Su HB (2004) Extracorporeal high intensity focused ultrasound in the treatment of patients with large hepatocellular carcinoma. Ann Surg Oncol, 1:1061–1069.

Kennedy JE, Wu F, ter Haar GR, Gleeson FV, Phillips RR, Middleton MR and Cranston D (2004) High-intensity focused ultrasound for the treatment of liver tumours. Ultrasonics, 42:931-935.

Cline HE, Schenck JF, Hynynen K, Watkins RD, Souza SP and Jolesz FA (1992) MR-guided focused ultrasound surgery. J Comput Assist Tomogr, 16:956-965.

Cline HE, Hynynen K, Watkins RD, Adams WJ, Schenck JF, Ettinger RH, Freund WR, Vetro JP and Jolesz FA (1995) Focused US system for MR imaging-guided tumor ablation. Radiology, 194:731-737.

Hynynen K, Vykhodtseva NI, Chung AH, Sorrentino V, Colucci V and Jolesz FA (1997) Thermal effects of focused ultrasound on the brain: determination with MR imaging. Radiology, 204:247-253.

Jolesz FA and Hynynen K (2008) Magnetic resonance image-guided focused ultrasound surgery. Cancer J, 8: S100-S112.

Okada A, Murakami T, Mikami K, Onishi H, Tanigawa N, Marukawa T and Nakamura H (2006) A case of hepatocellular carcinoma treated by MR-guided focused ultrasound ablation with respiratory gating. Magn Reson Med Sci, 5:167–171.

Kopelman D, Inbar Y, Hananel A, Freundlich D, Castel D, Perel A, Greenfeld A, Salamon T, Sareli M, Valeanu A and Papa M (2006) Magnetic resonance-guided focused ultrasound surgery (MRgFUS): ablation of liver tissue in a porcine model. Eur J Radiol, 59:157–162.

Chapelon JY, Margonari J, Theillère Y, Gorry F, Vernier F, Blanc E and Gelet A (1992) Effects of high-energy focused ultrasound on kidney tissue in the rat and the dog. Eur Urol, 22:147-152.

Hacker A, Michel MS, Marlinghaus E, Kohrmann KU and Alken P (2005) Clinical study with extracorporeal transducer ultrasound guidance 2005: Extracorporeally induced ablation of renal tissue by high-intensity focused ultrasound. BJU Int, 97:779–785.

Illing RO, Kennedy JE, Wu F, ter Haar GR, Protheroe AS, Friend PJ, Gleeson FV, Cranston DW, Phillips RR and Middleton MR (2005) The safety and feasibility of extracorporeal high-intensity focused ultrasound (HIFU) for the treatment of liver and kidney tumours in a Western population. Br J Cancer, 93:890-985.

Klingler HC, Susani M, Seip R, Mauermann J, Sanghvi N and Marberger MJ (2008) A novel approach to energy ablative therapy of small renal tumours: Laparoscopic high-intensity focused ultrasound. Eur Urol, 53:810–816.

Ries M, Denis de Senneville B, Roujol S, Berber Y, Quesson B and Moonen CTW (2010) Real-Time 3D Target Tracking in MRI-guided Focused Ultrasound Ablations in moving Tissues, Mag Reson Med, 64:1704-1712.

Quesson B, Laurent C, Maclair G, Denis de Senneville B, Mougenot C, Ries M, Carteret T, Rullier A and Moonen CTW (2011) Real-time volumetric MRI thermometry of focused ultrasound ablation in vivo: a feasibility study in pig liver and kidney. NMR Biomed, 24:145–153.

Denis de Senneville B, Quesson B and Moonen CTW (2005) Magnetic Resonance Temperature imaging. Int J Hyperthermia, 21:515-531.

Rieke V and Butts Pauly K (2008) MR Thermometry. J Magn Reson Imaging, 27:376–390.

Hynynen K, McDannold N, Clement G, Jolesz FA, Zadicario E, Killiany R, Moore T and Rosen D (2006) Pre-clinical testing of a phased array ultrasound system for MRI-guided noninvasive surgery of the brain - a primate study. Eur J Radiol, 59:149–156.

Gellermann J, Wlodarczyk W, Feussner A, Fahling H, Nadobny J, Hildebrandt B, Felix R and Wust P (2005) Methods and potentials of magnetic resonance imaging for monitoring radiofrequency hyperthermia in a hybrid system. Int J Hyperthermia, 21:497-513.

Denis de Senneville B, Mougenot C, Quesson B, Dragonu I, Grenier N and Moonen CTW (2007) MR-thermometry for monitoring tumor ablation. Eur Radiol, 17:2401-2410.

Sapareto SA and Dewey WC (1984) Thermal dose determination in cancer therapy. Int J Radiat Oncol Biol Phys, 10:787-800.

Mougenot C, Köhler MO, Enholm J, Quesson B and Moonen CTW (2011) Quantification of near-field heating during volumetric MR-HIFU ablation, Med Physics, 38:272-282.

Baron P, Ries M, Deckers R, de Greef M, Tanttu J, Köhler M, Viergever MA, Moonen CT and Bartels LW (2013) In vivo T2-based MR thermometry in adipose tissue layers for high-intensity focused ultrasound near-field monitoring, Magn Reson Med, 72:1057-1064.

Goss S, Frizzell L and Dunn F (1979) Ultrasonic absorption and attenuation in mammalian tissues. Ultrasound Med Biol, 5:181-186.

Liu HL, Chang H, Chen WS, Shih TC, Hsiao JK and Lin WL (2007) Feasibility of transrib focused ultrasound thermal ablation for liver tumors using a spherically curved 2D array: A numerical study. Med Physics, 349:3436-3448.

Bobkova S, Gavrilov L, Khoklova V, Shaw A and Hand J (2010) Focusing of high intensity ultrasound through the rib cage using a therapeutic random phased array. Ultrasound Med Biol, 366:888-906.

Liu H, McDannold N and Hynynen K (2005) Focal beam distortion and treatment planning in abdominal focused ultrasound surgery. Med Physics, 325:1270-1280.

Daum D, Smith N, King R and Hynynen K (1999) In vivo demonstration of noninvasive thermal surgery of the liver and kidney using an ultrasonic phased array. Ultrasound Med Biol, 257:1087-1098.

Wu F, Wang ZB, Cheng WZ, Zu H, Bai J, Zou JZ, Li KQ, Jin CB, Xie FL and Su HB (2004) Extracorporeal high intensity focused ultrasound ablation in the treatment of patients with large hepatocellular carcinoma. Ann Surg Oncol, 1112:1061-1069.

Li J, Xu G, Gu M, Luo G, Rong Z, Wu P and Xia J (2007) Complications of high intensity focused ultrasound in patients with recurrent and metastatic abdominal tumors. World J Astroenterol, 13:2747-2751.

Jung SE, Cho SH, Jang JH and Han JY (2011) High-intensity focused ultrasound ablation in hepatic and pancreatic cancer: complications. Abdom imaging, 362:185-195.

Ibbini MS, Ebbini ES and Cain CA (1990) NxN square-element ultrasound phased-array applicator: simulated temperature distributions associated with directly synthesized heating patterns. IEEE Trans Ultrason Ferroelectr Freq Control, 37:491-500.

McGough RJ, Kessler ML, Ebbini ES and Cain CA (1996) Treatment planning for hyperthermia with ultrasound phased arrays. IEEE Trans Ultrason Ferroelectr Freq Control, 43:1074–1084.

Botros YY, Volakis JL, VanBaren P and Ebbini ES (1997) A hybrid computational model for ultrasound phased-array heating in presence of strongly scattering obstacles. IEEE Trans Biomed Eng, 44:1039–1050.

Civale J, Clarke R, Rivens I and ter Haar G (2006) The use of a segmented transducer for rib sparing in HIFU treatments. Ultrasound Med Biol, 32:1753–1761.

Quesson B, Merle M, Kohler M, Mougenot C, Roujol S, Denis de Senneville B and Moonen CTW (2010) A method for MRI guidance of intercostal high intensity focused ultrasound ablation in the liver. Med Physics, 376:2533-2540.

De Greef M, Schubert G, Wijlemans JW, Koskela J, Bartels LW, Moonen CTW and Ries M (2014) Intercostal high intensity focused ultrasound for liver ablation: The influence of beam shaping on sonication efficacy and near-field risks, submitted to Med Physics.

Gélat P, Ter Haar G and Saffari N (2014) A comparison of methods for focusing the field of a HIFU array transducer through human ribs. Phys Med Biol, 59:3139-3171.

Aubry JF, Pernot M, Marquet F, Tanter M and Fink M (2008) Transcostal high-intensity-focused ultrasound: ex vivo adaptive focusing feasibility study. Phys Med Biol, 53:2937–2951.

Fink M (1997) Time reversed acoustics. Physics Today, 50:34–40.

Fink M, Montaldo G and Tanter M (2003) Time reversal acoustics in biomedical engineering, Ann Rev Biomed Eng, 5:465–497.

Tanter M, Pernot M, Aubry J-F, Montaldo G, Marquet F and Fink M (2007) Compensating for bone interfaces and respiratory motion in high intensity focused ultrasound. Int J Hyperthermia, 23:141–151.

Tanter M, Aubry JF, Gerber J, Thomas JL and Fink M (2001) Optimal focusing by spatiotemporal inverse filter. I. Basic principles. J Acoust Soc Am, 110:37-47.

Gélat P, Ter Haar G and Saffari N (2011) Modelling of the acoustic field of a multi-element HIFU array scattered by human ribs. Phys Med Biol, 56:5553-5581.

Cochard E, Prada C, Aubry JF and Fink M (2009) Ultrasonic focusing through the ribs using the DORT method. Med Physics, 36:3495-3503.

Cochard E, Aubry JF, Tanter M and Prada C (2011) Adaptive projection method applied to three-dimensional ultrasonic focusing and steering through the ribs. J Acoust Soc Am, 130:716-723.

Prada C (2002) Detection and imaging in complex media with the DORT method. Top Appl Phys, 84:107–133.

Prada C, Manneville S, Spoliansky D and Fink M (1996) Decomposition of the time reversal operator: detection and selective focusing on two scatterers. J Acoust Soc Am, 99:2067–2076.

Gélat P, ter Haar G and Saffari N (2012) The optimization of acoustic fields for ablative therapies of tumours in the upper abdomen. Phys Med Biol, 57:8471-8497.

Marquet F, Aubry J, Pernot M, Fink M and Tanter M (2011) Optimal transcostal high-intensity focused ultrasound with combined real-time 3D movement tracking and correction. Phys Med Biol, 56:7061-7068.

Hwang JH, Tu J, Brayman AA, Matula TJ and Crum LA (2006) Correlation between inertial cavitation dose and endothelial cell damage in vivo. Ultrasound Med Biol, 32:1611-1619.

Miller DL (2007) Overview of experimental studies of biological effects of medical ultrasound caused by gas body activation and inertial cavitation. Prog Biophys Mol Biol, 9:314-330.

Mirabell R, Nouet P, Rouzaud M, Bardina A, Heijira N and Schneider D (1998) Radiotherapy of bladder cancer: relevance of bladder volume changes in planning boost treatment. Int J Radiat Oncol Biol Phys, 41:741–746.

Langen KM, Willoughby TR, Meeks SL, Santhanam A, Cunningham A, Levine L and Kupelian PA (2008) Observations on real-time prostate gland motion using electromagnetic tracking. Int J Radiat Oncol Biol Phys, 71:1084–1090.

Smitmans MHP, Pos FJ, de Bois J, Heemsbergen WD, Sonke J-J, Lebesque JV and van Herk M. (2008) The influence of a dietary protocol on a cone beam CT-guided radiotherapy for prostate cancer patients. Int J Radiat Oncol Biol Phys, 71:1279–1286.

Emmott J, Sanghera B, Chambers J and Wong WL (2008) The effects of n-butylscopolamine on bowel uptake: an F-FDG PET study. Nuc Med Commun, 29:11–16.

Verhey LJ (1995) Immobilizing and positioning patients for radiotherapy. Sem Radiat Oncol, 5:100-114.

Zhang L, Chen WZ, Liu YJ, Hu X, Zhou K, Chen S, Peng L, Zhu H, Zou HL, Bai J and Wang ZB (2010) Feasibility of magnetic resonance imaging-guided high intensity focused ultrasound therapy for ablating uterine fibroids in patients with bowel lies anterior to uterus. Eur J Radiol, 73:396-403.

Denis de Senneville B, Ries M, Maclair G and Moonen CTW (2011) MR-guided thermotherapy of abdominal organs using a robust PCA-based motion descriptor. IEEE Trans Med Imaging, 30:1987-1995.

Denis de Senneville B, El Hamidi A, Moonen C (2014) A direct PCA-based approach for real-time description of physiological organ deformations, IEEE Transactions on Medical Imaging, 10.1109/TMI.2014.2371995

Gedroyc WM (2006) New clinical applications of magnetic resonance-guided focused ultrasound. Top Magn Reson Imaging, 17:189–194.

Holbrook AB, Ghanouni P, Santos JM, Dumoulin C, Medan Y and Pauly KB (2014) Respiration based steering for high intensity focused ultrasound liver ablation. Mag Reson Med, 71:797–806.

Cornelis F, Grenier N, Moonen CT and Quesson B (2011) In vivo characterization of tissue thermal properties of the kidney during local hyperthermia induced by MR-guided high-intensity focused ultrasound. NMR Biomed, 24:799-806.

Wijlemans JW, de Greef M, Schubert G, Bartels LW, Moonen CTW, van den Bosch MAAJ and Ries M (2014) A clinically feasible treatment protocol for MR-guided HIFU ablation in the liver. Invest Radiol, In press.

Wijlemans JW, van Breugel JMM, de Greef M, Moonen CTW, van den Bosch MAAJ and Ries M (2014) Spontaneous breathing vs mechanical ventilation for respiratory-gated MR-HIFU ablation in the liver, Focused Ultrasound 2014, 4th International Symposium.

Auboiroux V, Petrusca L, Viallon M, Muller A, Terraz S, Breguet R, Montet X, Becker CD and Salomir R (2014) Respiratory-Gated MRgHIFU in Upper Abdomen Using an MR-Compatible In-Bore Digital Camera. BioMed Res Int, 421726.

Koh TS, Thng CH, Lee PS, Hartono S, Rumpel H, Goh BC and Bisdas S (2008) Hepatic metastases: in vivo assessment of perfusion parameters at dynamic contrast-enhanced MR imaging with dual-input two-compartment tracer kinetics model. Radiol, 249:307-320.

de Zwart JA, Vimeux FC, Palussiere J, Salomir R, Quesson B, Delalande C and Moonen CTW (2001) On-line correction and visualization of motion during MRI-controlled hyperthermia. Mag Reson Med, 45:128–137.

Sotiras A, Davatzikos C and Paragios N (2013) Deformable medical image registration: a survey. IEEE Trans Med Imaging, 32:1153–1190.

Barron JL, Fleet DJ and Beauchemin SS (1994) Performance of optical flow techniques. Int J Comput Vis, 12:43–77.

Denis de Senneville B, Mougenot C and Moonen CTW (2007) Real-time adaptive methods for treatment of mobile organs by MRI-controlled high-intensity focused ultrasound. Mag Reson Med, 57:319–330.

Roujol S, Ries M, Quesson B, Moonen CTW and Denis de Senneville B (2010) Real-time MR-thermometry and dosimetry for interventional guidance on abdominal organs. Mag Reson Med, 63:1080–1087.

Roujol S, Ries M, Moonen CTW and Denis de Senneville B (2011) Automatic nonrigid calibration of image registration for real time MR-guided HIFU ablations of mobile organs. IEEE Trans Med Imaging, 30:1737–1745.

Graham SJ, Bronskill MJ and Henkelman RM (1998) Time and temperature dependence of MR parameters during thermal coagulation of ex vivo rabbit muscle. Mag Reson Med, 39:198–203.

Köhler M, Denis de Senneville B, Quesson B, Moonen CTW and Ries M (2011) Spectrally selective pencil-beam navigator for motion compensation of MR-guided high-intensity focused ultrasound therapy of abdominal organs. Mag Reson Med, 66:102-111.

Günther M and Feinberg DA (2004) Ultrasound-guided MRI: Preliminary results using a motion phantom. Mag Reson Med, 52:27-32.

Feinberg DA, Giese D, Bongers DA, Ramanna S, Zaitsev M, Markl M and Günther M (2010) Hybrid ultrasound MRI for improved cardiac imaging and real-time respiration control. Mag Reson Med, 63:290-296.

Arnold P, Preiswerk F, Fasel B, Salomir R, Scheer K and Cattin PC (2011) 3D organ motion prediction for MR-guided high intensity focused ultrasound. Med Image Comput Comput Assist Interv, 14:623-630.

Brix L, Ringgaard S, Sorensen TS and Poulsen PR (2014) Three-dimensional liver motion tracking using real-time two-dimensional MRI. Med Physics, 41:042302.

Stemkens B, Tijssen R, Denis de Senneville B, Lagendijk J and van den Berg C (2014) Retrospective Reconstruction of 3D Radial MRI Data to Evaluate the Effect of Abdominal Compression On 4D Abdominal Organ Motion. Med Physics, 41:538-538.

Stemkens B, Tijssen R, Denis de Senneville B, Heerkens HD, van Vulpen LJ, Lagendijk J, and van den Berg C (2014) Optimizing 4D-MRI data sampling for respiratory motion analysis of pancreatic tumors, International Journal of Radiation Oncology, Biology, Physics, 10.1016/j.ijrobp.2014.10.050.

Pernot M, Aubry JF, Tanter M, Thomas JL and Fink M (2003) High power transcranial beam steering for ultrasonic brain therapy. Phys Med Biol, 48:2577-2589.

Lorenzo de Oliveira P, Denis de Senneville B, Dragonu I and Moonen CTW (2010) Rapid motion correction in MR-guided high-intensity focused ultrasound heating using real-time ultrasound echo information, NMR Biomed, 23:1103–1108.

Ries M, Denis de Senneville B, Regard Y and Moonen CTW (2012) Combined magnetic resonance imaging and ultrasound echography guidance for motion compensated HIFU interventions, 12th International Symposium on Therapeutic Ultrasound: AIP Publishing, 202–206.

Auboiroux V, Petrusca L, Viallon M, Goget T, Becker CD and Salomir R (2012) Ultrasonography-based 2D motion-compensated HIFU sonication integrated with reference-free MR temperature monitoring: a feasibility study ex vivo. Phys Med Biol, 57:N159-N171.

Denis de Senneville B, Regard Y, Moonen CTW and Ries M (2014) Combined Magnetic Resonance and Ultrasound Echography Guidance for direct and indirect motion tracking, 11th IEEE Int Symp on Biomedical Imaging (ISBI), Beijing.

Sapareto SA and Dewey WC (1984) Thermal dose determination in cancer therapy. Int J Radiat Oncol Biol Phys, 10:787-800.

Peters RD and Henkelman RM (2000) Proton-resonance frequency shift MR thermometry is affected by changes in the electrical conductivity of tissue. Magn Reson Med, 43:62-71.

De Poorter J, De Wagter C, De Deene Y, Thomson C, Stahlberg F and Achten E (1994) The proton resonance frequency shift method compared with molecular diffusion for quantitative measurement of two dimensional time dependent temperature distribution in phantom. J Magn Reson Imaging, 103:234-241.

Young IR, Hajnal JV, Roberts IG, Ling JX, Hill-Cottingham RJ, Oatridge A and Wilson J A (1996) An evaluation of the effects of susceptibility changes on the water chemical shift method of temperature measurement in human peripheral muscle. Magn Reson Med, 36:366-374.

Morikawa S, Inubushi T, Kurumi Y, Naka S, Sato K, Demura K, Tani T and Haque HA (2004) Feasibility of respiratory triggering for MR-guided microwave ablation of liver tumors under general anesthesia. Cardiovas Interv Radiol, 27:370–373.

Maier F, Krafft AJ, Yung JP, Stafford RJ, Elliott A, Dillmann R, Semmler W and Bock M (2012) Velocity navigator for motion compensated thermometry. MAGMA, 25:15-22.

Weidensteiner C, Kerioui N, Quesson B, Denis de Senneville B, Trillaud H and Moonen CTW (2004) Stability of real-time MR temperature mapping in healthy and diseased human liver. J Magn Reson Imaging, 19:438–446.

Vigen KK, Daniel BL, Pauly JM and Butts K (2003) Triggered, navigated, multi-baseline method for proton resonance frequency temperature mapping with respiratory motion. Magn Reson Med, 50:1003-1010.

Hey S, Maclair G, Denis de Senneville B, Lepetit-Coiffe M, Berber Y, Köhler MO, Quesson B, Moonen CT and Ries M (2009) Online correction of respiratory-induced field disturbances for continuous MR-thermometry in the breast. Magn Reson Med, 61:1494-1499.

Rieke V, Vigen KK, Sommer G, Daniel BL, Pauly JM and Butts K (2004) Referenceless PRF shift thermometry. Magn Reson Med, 51:1223-1231.

McDannold N, Tempany CM, Jolesz FA and Hynynen K (2008) Evaluation of referenceless thermometry in MRI-guided focused ultrasound surgery of uterine fibroids. J Magn Reson Imaging, 228:1026-1032.

Holbrook AB, Santos JM, Kaye E, Rieke V and Pauly KB (2010) Real-time MR thermometry for monitoring HIFU ablations of the liver. Magn Reson Med, 63:365–373.

Kuroda K, Kokuryo D, Kumamoto E, Suzuki K, Matsuoka Y and Keserci B (2006) Optimization of self-reference thermometry using complex field estimation. Magn Reson Med, 56:835–843.

Zou C, Shen H, He M, Tie C, Chung YC and Liu X (2013) A fast referenceless PRFS-based MR thermometry by phase finite difference. Phys Med Biol, 58:5735-5751.

Shi C, Xie G, Song Y, Zou C, Liu X and Zhou S (2014) Referenceless PRFS MR Thermometry Using Partial Separability Model. Appl Magn Reson, 45:93-108.

Liu T, Khalidov I, de Rochefort L, Spincemaille P, Liu J, Tsiouris AJ and Wang Y (2011) A novel background field removal method for MRI using projection onto dipole fields (PDF). NMR Biomed, 24:1129–1136.

Schweser F, Lehr BW, Deistung A and Reichenbach JR (2010) A Novel Approach for Separation of Background Phase in SWI Phase Data Utilizing the Harmonic Function Mean Value Property, Proceedings to the 18th meeting of the International Society for Magnetic Resonance in Medicine, Stockholm.

Salomir R, Viallon M, Kickhefel A, Roland J, Morel DR, Petrusca L, Auboiroux V, Goget T, Terraz S, Becker CD and Gross P (2012) Reference-free PRFS MR-thermometry using near-harmonic 2-D reconstruction of the background phase. IEEE Trans Med Imaging, 31:287–301.

Grissom WA, Rieke V, Holbrook AB, Medan Y, Lustig M, Santos J, McConnell MV and Pauly KB (2010) Hybrid referenceless and multibaseline subtraction MR thermometry for monitoring thermal therapies in moving organs. Med Physics, 37:5014-5026.

Denis de Senneville B, Roujol S, Moonen CTW and Ries M (2010) Motion Correction in MR Thermometry of Abdominal Organs: A Comparison of the Referenceless vs the Multibaseline Approach. Magn Reson Med, 64:1373-1381.

Tsao J, Boesiger P and Pruessmann KP (2003) k-t BLAST and k-t SENSE: dynamic MRI with high frame rate exploiting spatiotemporal correlations. Magn Reson Med, 50:1031-1042.

Todd N, Prakash J, Odéen H, de Bever J, Payne A, Yalavarthy P and Parker DL (2014) Toward real-time availability of 3D temperature maps created with temporally constrained reconstruction. Magn Reson Med, 71:1394–1404.

Roujol S, Denis de Senneville B, Hey S, Moonen CTW and Ries M (2012) Robust adaptive extended Kalman filtering for real time MR-thermometry guided HIFU interventions. IEEE Trans Med Imaging, 31:533-542.

Denis de Senneville B, Roujol S, Hey S, Moonen CTW and Ries M (2013) Extended Kalman filtering for continuous volumetric MR-temperature imaging. IEEE Trans Med Imaging, 32:711-718.